\documentclass[aps,prl,preprint,showpacs]{revtex4}
\usepackage{mathrsfs}
\usepackage{graphicx}
\usepackage{amssymb}
\usepackage{amsmath}

\newcommand{\G}{ \mathscr{G} }
\usepackage{setspace}

\begin{document}

\title{Quantum Interference of Tunably Indistinguishable Photons from \\Remote Organic Molecules}

\author{R. Lettow$^{1}$, Y. L. A. Rezus$^{1}$, A. Renn$^{1}$, G. Zumofen$^{1}$, E. Ikonen$^{2}$, S. G\"otzinger$^{1}$, V. Sandoghdar$^{1}$}

\affiliation{$^1$Laboratory of Physical Chemistry and optETH, ETH
Zurich, CH-8093 Zurich, Switzerland \\ $^2$Metrology Research
Institute, Helsinki University of Technology (TKK) and \\ Centre for
Metrology and Accreditation (MIKES), P.O. Box 3000, FI-02015 TKK,
Finland}

\begin{abstract}
We demonstrate two-photon interference using two remote single
molecules as bright solid-state sources of indistinguishable
photons. By varying the transition frequency and spectral width of
one molecule, we tune and explore the effect of photon
distinguishability. We discuss future improvements on the brightness
of single-photon beams, their integration by large numbers on chips,
and the extension of our experimental scheme to coupling and
entanglement of distant molecules.
\end{abstract}


\maketitle

Recent developments in quantum engineering have redrawn the
attention of scientists to the phenomenon of interference between
single
photons~\cite{Santori:02,Legero:04,kiraz:05,Beugnon:06,Maunz:07,Sanaka:09,Bennett:09}
for its potential in applications such as entanglement
generation~\cite{Mandel:88a} and optical quantum
computing~\cite{Knill2001,Obrien:03}. In a two-photon quantum
interference (TPQI) experiment indistinguishable single photons from
two independent beams enter the two input ports of a 50-50 beam
splitter and leave together in one of the two output
ports~\cite{Paul:86}. Although the pioneering work on this topic
used photon pairs created in parameteric down
conversion~\cite{Hong:87}, ideally, it is desirable to use a large
number of bright independent sources of Fourier-limited photons. It
turns out that single quantum emitters are predestined for this task
because they are intrinsically small and can emit lifetime-limited
photons one at a time~\cite{Lounis:05}. Indeed, atoms and ions in
vacuum chambers have been successfully used in the context of TPQI
interference measurements~\cite{Legero:04,Beugnon:06,Maunz:07}.
However, atom reloading time, difficulties in efficient light
collection, and scaling to large numbers of emitters pose major
challenges to various realizations. Solid-state emitters such as
semiconductor quantum dots, color centers, and molecules are
scalable on chips, can have large emission rates, and lend
themselves to highly efficient collection schemes~\cite{Barnes:02}.
However, they face the main hurdles of spectral dephasing and
inhomogeneity, which make it difficult to find independent emitters
that generate indistinguishable photons. In this work, we show that
organic molecules embedded in organic matrices master all these
challenges and discuss the conditions for tolerating deviations from
the ideal case.

\begin{figure} [t]
\begin{center}
\includegraphics[width=7.5cm]{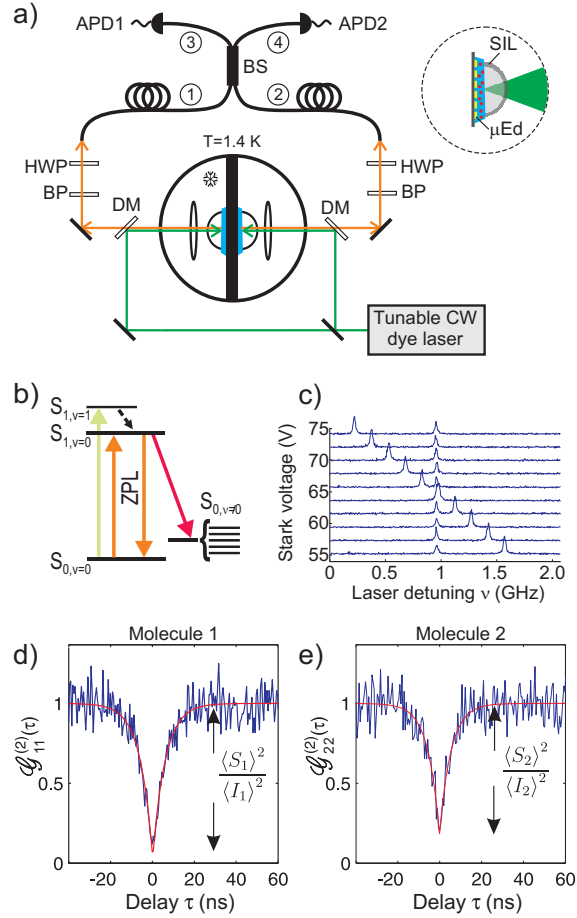}
\caption{a) Schematic diagram of the optical setup. DM, dichroic
mirror; BS, beam-splitter; BP, bandpass filter; HWP, half-wave
plate; APD, avalanche photodiode; SIL, solid-immersion lens;
$\mu$Ed, gold microelectrodes. b) Energy level diagram of a DBATT
molecule. c) Fluorescence excitation spectra of two selected
molecules as a function of the voltage on the microelectrodes of one
sample. This Stark effect was linear and well described by the
relation $\Delta\nu = 50 \times(V-63)$ (with $\Delta\nu$ in MHz and
V in Volts). d, e) Intensity autocorrelation functions of the two
molecules. See text for the details of the theoretical fit.
\label{scheme}}
\end{center}
\end{figure}

The organic dye molecules in this study were dibenzanthanthrene
(DBATT) embedded in {\em n}-tetradecane at a concentration of about
$10^{-6}$M. As illustrated in Fig.~\ref{scheme}a and described in
detail in Ref.~\cite{Lettow:07}, we used separate microscopes and
samples to extract indistinguishable photons on the zero-phonon
lines (ZPLs) of two remote molecules. At cryogenic temperatures
($T<$2~K), DBATT displays a sharp lifetime-limited ZPL
($\gamma_0\sim$17 MHz) between the ground vibrational level of the
electronic ground state $S_{\rm 0,v=0}$ and the ground vibrational
level of the electronic excited state $S_{\rm 1,v=0}$ at 589~nm (see
Fig.~\ref{scheme}b)~\cite{Brunel:99}. We used a narrow-band
($<1$~MHz) dye laser to address molecules across the inhomogeneous
spectral distribution of the sample ($\sim$2~THz)~\cite{Moerner:89}.
As the frequency of the laser was scanned, ZPLs of individual
molecules were excited selectively, and we recorded the
Stokes-shifted fluorescence on the $S_{\rm 1,v=0}\rightarrow S_{\rm
0,v\neq 0}$ transitions (see Fig.~\ref{scheme}b) to detect each
molecule~\cite{Orrit:90}. To obtain the same ZPL frequency for two
molecules in the two samples, we tuned the resonance of one molecule
by applying a voltage to the gold microelectrodes fabricated on its
substrate (see Fig.~\ref{scheme}c).

Once we had prepared two molecules with identical ZPLs, we generated
Fourier-limited single photons from them by tuning the dye laser
frequency to the transition between the ground state and the first
vibrational level of the electronic excited state ($S_{\rm 1,v=1}$).
We found that despite having the same ZPL, the $S_{\rm
0,v=0}\rightarrow S_{\rm 1,v=1}$ transition frequencies were
typically not the same for the two selected molecules. Nevertheless,
these transitions overlapped within their linewidths of about
30~GHz~\cite{Lettow:07}, allowing us to excite the two molecules
equally strongly by a suitable adjustment of the laser frequency.
The $S_{\rm 1,v=1}$ rapidly relaxes to the $S_{\rm 1,v=0}$ state
which has a lifetime of 9.5~ns determined by a radiative decay to
$S_{\rm 0,v=0}$ (ZPL) or $S_{\rm 0,v\neq 0}$ with a branching ratio
of about 0.5 (see Fig.~\ref{scheme}b). The emission on the ZPL with
a coherence length of about 3~m yielded up to one million counts on
the detector after passing a bandpass filter to reject the
excitation light and the Stokes-shifted
fluorescence~\cite{Lettow:07}. It is worth emphasizing that the
transition dipole associated with the ZPL has a well-defined
orientation with respect to the backbone of the molecular structure,
leading to a linearly polarized emission.

\begin{figure} [b]
\begin{center}
\includegraphics[width=7.3cm]{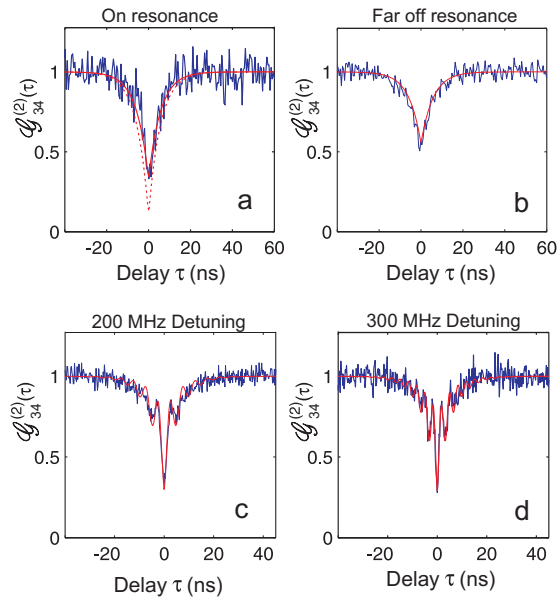}
\caption{a) Intensity cross correlation function of photons with the
same ZPL frequency (on-resonant) exiting the output ports of the
beam splitter. b) Same as (a) if the ZPL of one molecule is
frequency detuned by about 5~GHz (far off resonant). c,d) Same as
(a) if the two emitters are frequency detuned by 200~MHz and
300~MHz, respectively. The red curves are calculations based on
Eqn.~(\ref{EqnTPI}) with $\eta=0.5$ (solid) and $\eta=1$ (dashed),
respectively.\label{g2}}
\end{center}
\end{figure}

To realize an arrangement for a TPQI measurement, the ZPL emissions
of the two molecules were focussed into the two arms of a
single-mode polarization-maintaining fiber beam splitter, which
conveniently ensured spatial mode matching of the two beams (see
Fig.~\ref{scheme}a). Two half-wave plates were used to align the
input polarizations. The outputs of this device were directed to two
avalanche photodiodes (APDs) connected to a time-to-amplitude
converter that functioned in a start-stop configuration and allowed
us to record intensity correlations with a time resolution of
$\sim$800~ps. In a first step, we always verified that each beam
consisted of single photons by recording its intensity
autocorrelation function $g^{(2)}(\tau)$ separately when the other
fiber input was blocked. Figures~\ref{scheme}d,e display strong
photon antibunching at zero time delay $\tau$ for two molecules
(i=1,2) from the two samples. In the ideal case, one expects
$g^{(2)}_{\rm ii}(0)=0$~\cite{Loudon}. If the single-photon
intensity ($S_{\rm i}$) of each molecule is accompanied by an
uncorrelated background ($B_{\rm i}$), the overall detected
intensity $I_{\rm i}=S_{\rm i}+B_{\rm i}$ satisfies $\G_{\rm
ii}^{(2)}(\tau)= 1+\frac{\langle S_{\rm i} \rangle^2}{\langle I_{\rm
i} \rangle^2}[g_{\rm ii}^{(2)}(\tau)-1]$. In what follows, we use
the calligraphic notation $\G(\tau)$ to denote a measurement in the
presence of background and reserve $g(\tau)$ for the ideal
correlation functions. The measured autocorrelation functions
$\G^{(2)}_{\rm ii}(0)=1-\frac{\langle S_{\rm i} \rangle^2}{\langle
I_{\rm i} \rangle^2}$ shown in Figs.~\ref{scheme}d,e reveal that
$\frac{\langle S \rangle}{\langle I \rangle} \geq 90\%$. In our
experiment, the residual background stemmed from neighboring
nonresonant molecules in the excitation spot, which can be
eliminated by reducing the DBATT concentration during sample
preparation.

For a beam consisting of single photons from two different
molecules, $g^{(2)}(0)$ is reduced to 0.5 because the detection of a
photon from one molecule does not impede detecting a second one from
the other molecule~\cite{Loudon}. The fascinating feature of a TPQI
experiment is that even photons from two independent emitters can
yield a perfect anticorrelation. Here, two photons that are
indistinguishable in frequency, linewidth, spatial mode, and
polarization enter the two input arms of a beam splitter (labeled 1
and 2) and coalesce in one of the output ports (labeled 3 and 4) so
that the probability of simultaneously detecting one photon in each
arm vanishes at zero time delay~\cite{Legero:06}. The theoretical
expression for the intensity cross correlation function of the two
outgoing modes of the beam splitter reads [see Supplementary
Material],
\begin{eqnarray}
\G^{(2)}_{\text{34}}(\tau) &=& c_1^2 \G^{(2)}_{11}(\tau) + c_2^2
\G^{(2)}_{22}(\tau)
                     \nonumber\\ + &2c_1c_2&\left\{1-\eta
                     \frac{\left<S_1 \right>\left<S_2 \right>}{\left<I_1 \right>\left<I_2 \right>}
                     |g^{(1)}_{11}(\tau)||g^{(1)}_{22}(\tau)|\cos(\Delta\omega\tau)\right\}
                     .\nonumber\\\label{EqnTPI}
\end{eqnarray}
where $ c_i = I_i/(I_1 + I_2)$. The first and second terms represent
the intensity autocorrelations of the individual sources as measured
experimentally and presented in Figs.~\ref{scheme}d and
\ref{scheme}e, whereas the term in brackets originates from the
mixed products of the two input intensities at frequency detuning
$\Delta\omega$. For a quantum emitter i, the intensity (second
order) and field (first order) autocorrelation functions are related
according to $g_{\rm ii}^{(2)}(\tau)=1-|g_{\rm ii}^{(1)}(\tau)|^2$
where $g^{(1)}_{\text{ii}}(\tau)=e^{-i\omega_{\rm
i}\tau}e^{-\gamma{|\tau|/2}}$ and $\gamma$ is the homogeneous
linewidth of the emitter~\cite{Loudon}. Thus, the measurements of
$\G_{\rm ii}^{(2)}$ determine both the ratios $\frac{\langle S_{\rm
i} \rangle}{\langle I_{\rm i} \rangle}$ and $g^{(1)}_{\rm ii}$,
therefore fully characterizing $\G^{(2)}_{\text{34}}(\tau)$. In
practice, the visibility of the two-photon interference could be
reduced by factors other than the background light. We have,
therefore, included the phenomenological parameter $0 \leq \eta \leq
1$ in Eqn.~(\ref{EqnTPI}) to account for this effect.

The blue trace in Fig.~\ref{g2}a displays
$\G^{(2)}_{\text{34}}(\tau)$ for the two output ports of the beam
splitter when the inputs were photons emitted by the same two
molecules presented in Figs.~\ref{scheme}d,e. The fact that
$\G^{(2)}_{\text{34}}(0)<0.5$ is a clear proof of a quantum
interference and has its origin in the corpuscular nature of
single-emitter radiation~\cite{Paul:86,Legero:06}. The red curve
shows a very good agreement with the predictions of
Eqn.~(\ref{EqnTPI}) based on the data from the experimental
measurements of $\G^{(2)}_{\text{ii}}(\tau)$ and with the assumption
that $\eta = 0.5$. The origin of the contrast reduction was due to
polarization ellipticities caused by a number of depolarizing
elements such as dielectric mirrors and cryostat windows. For
comparison, the dashed red curve displays the prediction of the
calculations for $\eta = 1$.

To examine the impact of photon distinguishability on the cross
correlation function, we exploited the frequency tunability of our
emitters and changed the frequency of one molecule. First, we
explored the case of far-off detuning by setting $\Delta \omega/2\pi
\sim 5$~GHz. Figure~\ref{g2}b confirms that in this case, we
obtained $\G_{\text{34}}^{(2)}(0) \simeq 0.5$. The red curve shows a
very good agreement with the outcome of calculation with no free
parameters and assuming $\eta=0$. The latter condition is justified
by the fact that for photons with a large frequency difference, the
term proportional to the cosine in Eqn.~(\ref{EqnTPI}) is washed out
due to the limited time resolution of our detectors. At the same
time, this term suggests that a frequency mismatch between the two
emitters should introduce a time-dependent beat signal in the
coincidence counts~\cite{Legero:04}. Figures~\ref{g2}c and d display
the measured two-photon interference signal when the ZPL of one
molecule was detuned by $\Delta\omega/2\pi=$~200 MHz and
$\Delta\omega/2\pi=$~300 MHz, respectively. The solid red curves
show that again calculations based on $\eta=0.5$ provide excellent
agreement with the experimental findings. It is noteworthy that
although the two photons are clearly distinguishable in frequency,
we still find $\G_{\text{34}}^{(2)}(0)< 0.5$. Quantum beat signals
shown in Figs.~\ref{g2}c,d were observed for the interference of
delayed photons from a single atom~\cite{Legero:04}, but to our
knowledge, this is the first demonstration for photons from
independent sources.

\begin{figure}
\begin{center}
\includegraphics[width=7.5cm]{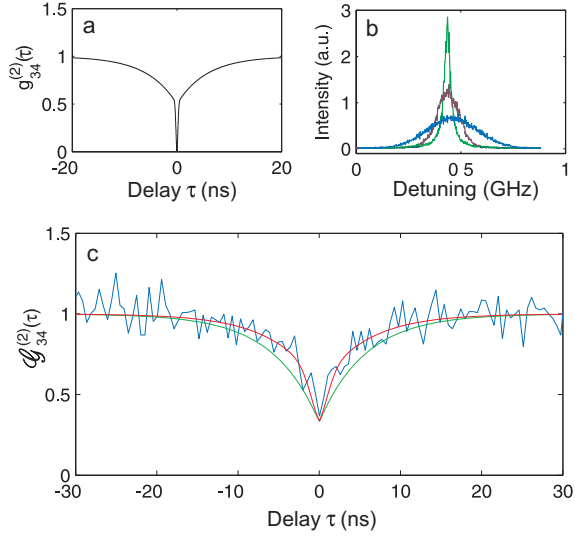}
\caption{a) Theoretical prediction of the $g_{34}^{(2)}$ function
when one molecule is broadened by about 120 times the natural
linewidth $\gamma_0$. b) Fluorescence excitation spectra of the ZPL
for different noise amplitudes applied to the electrode. FWHM=30~MHz
(blue), 120 (green), and 300 (red). c) Experimental measurement of
the two-photon interference where the ZPL of one molecule was
artificially broadened. The red curve is the prediction of
Eqn.~(\ref{EqnTPI}) for $\eta=0.5$ and the green curve displays the
data from Fig.~\ref{g2}a. \label{dephase} }
\end{center}
\end{figure}

A key asset of organic molecules is that they routinely exhibit
resonances with natural linewidth~\cite{Lounis:05}. Although other
solid-state systems such as semiconductor quantum dots and color
centers can, in principle, also reach this level of coherence, their
performance has been critically dependent on the sample quality and
reports of Fourier-limited emission from these systems are still
missing. Dephasing and spectral diffusion processes in solid-state
emitters result in fast variations of the emitter resonance
frequency and therefore fluctuations of $\Delta \omega$ in
Eqn~(\ref{EqnTPI}). As an example, Fig.~\ref{dephase}a shows the
theoretical predictions if one of the ZPL of one molecule were
broadened to a full width at half-maximum (FWHM) of 2~GHz. Although
we still find $g_{\text{34}}^{(2)}(0)=0$, a finite detector time
resolution of the order of 1~ns would wash out the contrast. In
order to explore this experimentally, we artificially broadened the
ZPL of one of the two molecules by applying quasi-white noise to its
sample microelectrodes. Figure~\ref{dephase}b illustrates the
resulting ZPL with FWHM=300~MHz ($\sim 18\gamma_0$) at the maximum
amplitude of the applied noise. We note in passing that the line
profiles are no longer Lorentzian because of the limited modulation
bandwidth of 1~MHz that was used in the Stark broadening process.
The blue trace in Fig.~\ref{dephase}c displays the resulting
$\G_{\text{34}}^{(2)}(\tau)$ measurement, and the red solid curve
shows the prediction of Eqn.~(\ref{EqnTPI}) with $\eta=0.5$. As
compared to the green curve which recasts the data of
Fig.~\ref{g2}a, the TPQI dip narrows but the contrast reduction is
not substantial because as in the case presented in Fig.~\ref{g2}d,
the response of our detectors has been sufficiently fast for
resolving the features of $\G_{\text{34}}^{(2)}$. In summary,
current photodetectors allow a considerable deviation of the
spectral coherence from the ideal Fourier-limited condition without
compromising the signature of the two-photon interference.
Nevertheless, it has to be born in mind that any dephasing or
spectral diffusion process reduces the probability of two-photon
coalescence after the beam splitter because it lowers the coherence
time of the photons ($1/\gamma$) with respect to their radiative
lifetime ($1/\gamma_0$).

Other important and desirable features of single-photon sources for
exploiting quantum interference measurements are high emission
rates, large collection efficiency, and integration of a large
number of sources. Solid-state systems and in particular organic
molecules promise to address all of these criteria at the same time.
By excitation to the $S_{\rm 1,v=1}$ state via short pulses, DBATT
can emit one photon per excited-state lifetime of 9.5~ns, thus
reaching a rate of few tens of MHz~\cite{Ahtee:09}. For an emitter
at the interface of a medium with a high refractive index (see
Fig.~\ref{scheme}a), collection efficiencies beyond 90\% can be
achieved by optimizing the choices of the solid-immersion lens and
the numerical aperture of the collecting lens~\cite{Koyama:99}. A
collection efficiency of about 50\% and an overall detection
efficiency (filters, detector quantum efficiency, etc.) of 10\%
would yield more than $10^7\times (0.5 \times 0.1)^2>10000$
coincidences per second. Furthermore, one can integrate a large
number of small solid-immersion lenses and independently addressable
microelectrodes on the sample to extract many single photon beams
simultaneously~\cite{Ahtee:09}.

Manipulation of Fourier-limited photons emitted by organic molecules
paves the way towards a number of interesting experiments. First,
our experimental setup can be readily used to perform spectroscopy
on one molecule with tunable single photons emitted by the second
molecule~\cite{Wrigge:08}. Furthermore, the two-photon interference
arrangement gives access to a conditional
entanglement~\cite{Moehring:07} of distant molecules. Although this
entanglement only lasts during the lifetime of the electronic state
(about ten nanoseconds), application of ultrafast pulses can allow a
large number of coherent qubit rotations~\cite{Gerhardt:09}.
Moreover, one can envision replacing free-space photon channels used
in our current experiment with on-chip dielectric
waveguides~\cite{Quan:09}. Such a photonic circuit would offer a
``hard-wire" compact network of many individually-addressable
single-photon sources for complex quantum information processing
tasks.

We thank M. Pototschnig and J. Hwang for experimental help and R.
Pfab and V. Ahtee for contribution to the initial phase of the
experiment. This work was supported by the Swiss National Science
Foundation (SNF) and ETH Zurich (Fellowship to Y.R. and the QSIT
project grant Nr.~PP-01 07-02). E.I. acknowledges the warm
hospitality during his numerous visits at ETH and grant Nr. 129971
from the Academy of Finland.

\end{document}